\documentclass[journal=jacsat,manuscript=article]{achemso}

\usepackage[version=3]{mhchem} 
\usepackage{epspdfconversion}
\usepackage{graphicx}
\usepackage{dcolumn}
\usepackage{bm}
\usepackage{subcaption}
\usepackage{hyperref}
\usepackage[T1]{fontenc}
\usepackage{array,physics}
\usepackage{multirow,tabularx}
\usepackage{booktabs}
\usepackage[flushleft]{threeparttable}

\author{Lokanath Patra}
\author{ M. R. Ashwin Kishore}
\affiliation[Central University of Tamil Nadu]
{Department of Physics, Central University of Tamil Nadu, Thiruvarur 610101, India}
\author{R. Vidya}
\affiliation{Department of Medical Physics, Anna University, Chennai 600025, India}
\author{Anja O. Sj{\aa}stad}
\author{H. Fjellv{\aa}g}
\affiliation[University of Oslo]
{Center for Materials Science and Nanotechnology and Department of Chemistry, University of Oslo, Box 1033 Blindern, N-0315 Oslo, Norway}
\author{P. Ravindran}
\affiliation[Central University of Tamil Nadu]
{Department of Physics, Central University of Tamil Nadu, Thiruvarur 610101, India}
\alsoaffiliation[University of Oslo]
{Center for Materials Science and Nanotechnology and Department of Chemistry, University of Oslo, Box 1033 Blindern, N-0315 Oslo, Norway}
\email{raviphy@cutn.ac.in}
\phone{+91-94890-54267}

\title[An \textsf{achemso} demo]
  {Electronic and magnetic structures of hole doped trilayer La${_{4-x}}$Sr${_x}$Ni${_3}$O${_8}$ from first principles calculations}

\abbreviations{IR,NMR,UV}

\begin{document}


\begin{abstract}
 The magnetic and electronic properties of trilayer La${_4}$Ni${_3}$O${_8}$, similar to hole doped cuprates, are investigated by performing full$-$potential linearized augmented plane wave method$-$based spin$-$polarized calculations with LDA and GGA functionals including Hubbard \textit{U} parameters to account for strong correlation effects. Based on these calculations, we found that La${_4}$Ni${_3}$O${_8}$ is a \textit{C}$-$type antiferromagnetic (\textit{C}$-$AFM) Mott insulator in agreement with previous experimental and theoretical observations.  Our calculations suggest that Ni$^{1+}$ and Ni$^{2+}$ ions are found to be in high spin state with an average valency of +1.33. Intermediate band gap states are originated from $d_{z^2}$ electrons of both types of Ni ions after including the strong correlation effects. In order to understand the role of hole doping on electronic structure, phase stability, and magnetic properties of La${_4}$Ni${_3}$O${_8}$, similar calculations have been performed for La${_{4-x}}$Sr${_x}$Ni${_3}$O${_8}$ as a function of $x$, using the supercell approach. We have found that the hole doping brings insulator$-$to$-$metal transition without changing the \textit{C}$-$AFM ordering though the magnetic moment is enhanced at both Ni sites. Moreover, these Ni atoms are always in an average valence state irrespective of hole doping or volume change. So the electronic properties of hole doped La${_4}$Ni${_3}$O${_8}$ can not be compared with hole doped cuprates which are high $T_C$.
\end{abstract}

\section{Introduction}
The discovery of superconductivity in cuprates~\cite{Bednorz1986} has aroused extended interest in this fascinating phenomenon in other transition metal oxides, especially in neighboring nickelates. Nickelates have the potential to exhibit physics similar to high$-$\textit{T$_C$} cuprates~\cite{Anisimov1999, Poltavets2007, Poltavets2009a} because the electronic configuration of constituent  Ni$^{+}$ (3\textit{d$^9$}) which is similar to Cu$^{2+}$ in the high$-$\textit{T$_C$} cuprates. Since these compounds appear to have electrons rather than holes as charge carriers, there has been considerable interest in the electronic properties of these materials. Furthermore, theoretical studies~\cite{Anisimov1999} suggest that only nickelates with Ni$^{+}$ (3\textit{d$^9$}, S = $\frac{1}{2}$) can form an antiferromagnetic (AFM) insulator which can be doped with Ni$^{2+}$ holes similar to hole doped superconducting cuprates.
\par
The nickelate La${_4}$Ni${_3}$O${_8}$ was discovered in the late 90's which contains Ni ions in a square$-$planar coordination, has become a focus of interest. It is a member of the so-called T$^{\prime}$ type Ln$_{n+1}$ Ni$_n$O$_{2n+2}$ (Ln = La, Nd; $n$ = 2, 3, and $\infty$) homologous series which can be prepared by low temperature reduction of the Ruddlesden$-$Popper (RP) Ln$_{n+1}$ Ni$_n$O$_{3n+1}$.~\cite{Poltavets2007a} La$-$Ni$-$O system was initially investigated by Bednorz and M{\"{u}}ller for the possibility of superconductivity with Ni in 2+ and 3+ oxidation states.~\cite{Bednorz1987} Their discovery stimulated the research interest to look for superconductivity in other transition metal oxides with low dimensional crystal structures like cuprates, which are favorable for superconductivity. Nickel in 1+ state in square planar coordination can be very similar to the high \textit{T$_C$} cuprates, but Ni with 1+ oxidation state was difficult to stabilize at that time. Anisimov \textit{et al.}~\cite{Anisimov1999} predicted theoretically that nickelates  having the mixed valence of Ni$^{+}$/Ni$^{2+}$ have electronic configuration similar to the hole doped superconducting cuprates Cu$^{2+}$/Cu$^{3+}$ which shows high temperature superconductivity.
\par
At low temperatures La$-$Ni$-$O system undergoes metal$-$to$-$insulator transition~\cite{Torrance1992} as a function of its thickness. This happens due to electron correlation, band effect and also by spin, orbital and charge ordering~\cite{Liu2014, Elbio2005, Blanca-Romero2011}. Pardo \textit{et al.}~\cite{Pardo2012} anticipated that the pressure induced metal to insulator transition takes place at zero temperature and high$-$spin state to low$-$spin state transition at 105\,K. Also La${_4}$Ni${_3}$O${_8}$ can either be in ferromagnetic$-$metallic low spin (LS) phase or an antiferromagnetic$-$insulating high spin (HS) phase. Dimensionality induced insulator to metal transition was reported by Liu \textit{et al.}~\cite{Liu2014} in La$_{n+1}$Ni$_n$O$_{2n+2}$ system as its dimensionality changes with variation with n ($n$ = 2, 3 and $\infty$). The magnetic properties of this material series are determined by the spin states of Ni$^{+}$ and Ni$^{2+}$. Correspondingly, Poltavets \textit{et al.}~\cite{Poltavets2010} and N ApRoberts-Warren\textit{et al.}~\cite{aproberts2011critical} shown from nuclear magnetic resonance (NMR) measurements that La${_4}$Ni${_3}$O${_8}$ undergoes an AFM transition at \textit{T$_N$}=105\,K. Pardo \textit{et al.}~\cite{Pardo2012, Pardo2010} and Liu \textit{et al.}~\cite{Liu2012} also reported that La${_4}$Ni${_3}$O${_8}$ is  AFM with \textit{C}$-$AFM ordering in the ground state, which is a molecular correlated insulator. Moreover, they have reported that  the insulating state is originating from strong interlayer coupling and electronic correlation effects.
\par
Using the muffin$-$tin orbital (MTO) based linear MTO (LMTO) and N$^{th}-$order MTO (NMTO)~\cite{Andersen1984, Andersen2000} methods, Sarkar \textit{et al.}~\cite{Sarkar2011} showed that La${_4}$Ni${_3}$O${_8}$  in the \textit{A}$-$AFM and FM ordering possess Ni in low$-$spin state, and correlated metallic ground state, which is against the experimental observations.~\cite{Poltavets2010} Recently, the electrical resistivity of La${_4}$Ni${_3}$O${_8}$ has been measured under pressure and a suppression of the high$-$spin state was observed.~\cite{Cheng2012} Nevertheless, neither metallic state nor superconductivity has been seen under pressure ($\sim$ 50 GPa) and this may be due to the displacement of the oxygen atoms at high pressures. Among various scarcely metallic 3\textit{d} electron systems, hole$-$doped nickel oxides with K$_2$NiF$_4$ type structure, e.g., (La, Sr)$_2$NiO$_4$, are found to prompt metallic conductivity, yet more gradually than in the cuprates.~\cite{Cava1991} Essentially more Sr substitution is important to prompt metallic$-$type conductivities i.e. the  La${_{2-x}}$Sr${_x}$NiO${_4}$ system shows low temperature conductivity for $x$ close to 1. Doping at $A$ site by a suitable ion is one of the tools which could induce tremendous variation in the electrical and magnetic behavior of these RP phases. The insulating La$_2$CuO$_4$ transforms to superconducting~\cite{Takagi1992} state when it is doped with holes i.e. some of the lanthanum atoms are replaced by bivalent alkaline earth ion or basic monovalent cation with suitable ionic size. Enthalpy of formation of La$_{2-x}$A$_x$CuO$_4$ ($A$  = Ba, Sr, Ca and Pb) indicates  that more basic $A-$site cations tend to energetically balance out higher oxidation states of $B-$site cations, which is responsible for the change in electrical properties in the parent material.~\cite{Dicarlo1992} To observe the change in properties as a function of doping at $A-$site, La${_{2-x}}$Sr${_x}$NiO${_4}$ is another interesting system to study. This material behaves like an AF$-$insulator for $x$=0. For the higher Sr substitution range i.e. 0.5 $\leq$ $x$ $\leq$ 0.9, La${_{2-x}}$Sr${_x}$NiO${_4}$ behaves like a non$-$metal and shows the electrical conduction with variable ranges of hopping mechanism at low temperature and exponentially activated conduction at comparatively higher temperatures. The metallic conductivity sets in for concentration of $x$ between 1.0 and 1.3. Resistivity data in the temperature range up to 25\,K follows T$^{1/2}$ dependence for the $x$ value beyond 1.3, where strong electron$-$electron interactions occur in the presence of disorder in a diffusive metallic phase.~\cite{K1994} Orthorhombic Nd$_2$NiO$_{4\pm\delta}$ undergoes a phase transition to pseudo$-$tetragonal symmetry upon substitution with Sr${^{2+}}$ at Nd site.~\cite{Greenblatt1990} The increased magnetic moment in the Nd$_{2-x}$Sr$_x$NiO$_4$ system with increase in $x$ is attributed to the frustration of AF pairing that originates from the mixed valence of Ni$^{2+}$/Ni$^{3+}$, while it showed a metal$-$semiconductor transition which is related to the onset of a charge density wave characteristic of half$-$filled conduction bands.~\cite{TakedaY1992}
\par
Kharton, \textit{et al.}~\cite{Kharton1999}  studied La$_2$Ni$_{1-x}$Fe$_x$O$_{4\pm\delta}$ ($x$ = 0.02 and 0.10) and La$_2$Ni$_{0.88}$Fe$_{0.02}$Cu$_{0.10}$O$_{4+\delta}$. Their studies showed that the electrical conductivity for these compositions increased with the substitution of iron, but the electrical conductivity slightly decreased with copper substitution. La$_{2-x}$$A$$_x$NiO$_4$ ($A$ = Ca, Sr and Ba) system has been investigated by Tang, \textit{et al.}~\cite{Tang2000} and they have reported that the solubility limit was  0 $\leq$ $x$ $\leq$ 0.6 for Ca, 0 $\leq$ $x$ $\leq$ 1.5 for Sr and 0 $\leq$ $x$ $\leq$ 1.1 for Ba. They have also found that both La$_{2-x}$Ca$_x$NiO$_4$ and La$_{2-x}$Ba$_x$NiO$_4$ compositions exhibited semiconducting behavior for all values of $x$. Moreover the La$_{2-x}$Sr${_x}$NiO${_4}$ compositions for $x$ \textless 0.6 were reported to be semiconductor from room temperature up to 527$^{\circ}$\,C. On the other hand, the compositions with $x$ \textgreater 0.6 showed a transition temperature from semiconductor$-$to$-$metal, in which the transition temperature decreased as $x$ increased.  For $x$ \textgreater 1.3, there is no such transition was identified from room temperature up to 527$^{\circ}$\,C. For $n$ = 2 RP, the substitution of lanthanum by calcium on La$_{3-x}$Ca${_x}$Ni$_2$O$_{7-\delta}$ was reported by Nedilko, \textit{et al.}.~\cite{Nedilko2004} A single phase has been obtained in the compositions for 0 $\leq$ $x$ $\leq$ 0.8. In addition, Ca dopant reduced the mean oxidation state of nickel ion from 2.63 ($x$ = 0) to 2.15 ($x$ = 0.8). Mogni, \textit{et al.}~\cite{MOGNI2006} reported that the electrical conductivity of the La$_{0.3}$Sr$_{2.7}$FeNiO$_{7-\delta}$ sample is higher than that of the La$_{0.3}$Sr$_{2.7}$Fe$_2$O$_{7-\delta}$ phase. The electrical conductivity of cobalt doped on Ni$-$site of La$_4$Ni$_{3-x}$Co$_x$O$_{10\pm\delta}$ (0.0 $\leq$ $x$ $\leq$ 3.0) was measured by Amow \textit{et al.}.~\cite{Amow2006} They have shown that the overall conductivity decreased as $x$ increased up to $x$ = 2.0. Conversely, the conductivity is found to  increase with the higher amount of Co, i.e. $x$ > 2.0. Marius Uv Nagell \textit{et al.}~\cite{nagell2015structural} have investigated the structural and magnetic properties of La$_4$(Co$_{1-x}$Ni$_x$)$_3$O$_{10+\delta}$ (0 $\leq$ $x$ $\leq$ 1). They found the structure to be monoclinic with $P2_1/a$ symmetry. La$_4$Co$_3$O$_{10}$ was found to be as an antiferromagnetic semiconductor where La$_4$Ni$_3$O$_{10}$ as a paramagnetic metal. Additionally, after fine grinding La$_4$Co$_3$O$_{10}$ showed ferromagnetic like features below 80\,K. Several experimental as well as theoretical studies have been  made so far on the electrical and magnetic properties of La${_4}$Ni${_3}$O${_8}$. However, the effect of dopant on the La site of the system has hitherto not been reported. As one can expect exotic behavior such as valence state transition, magnetic transitions, insulator$-$to$-$metal transitions etc by electron/hole doping, we have investigated the role of hole doping  on crystal structure, electronic structure, and magnetic properties of La${_4}$Ni${_3}$O${_8}$ using state-of-the-art density functional theoretical calculations.
\section{Structural aspects}
\begin{figure}[!t]
\includegraphics[scale=0.4]{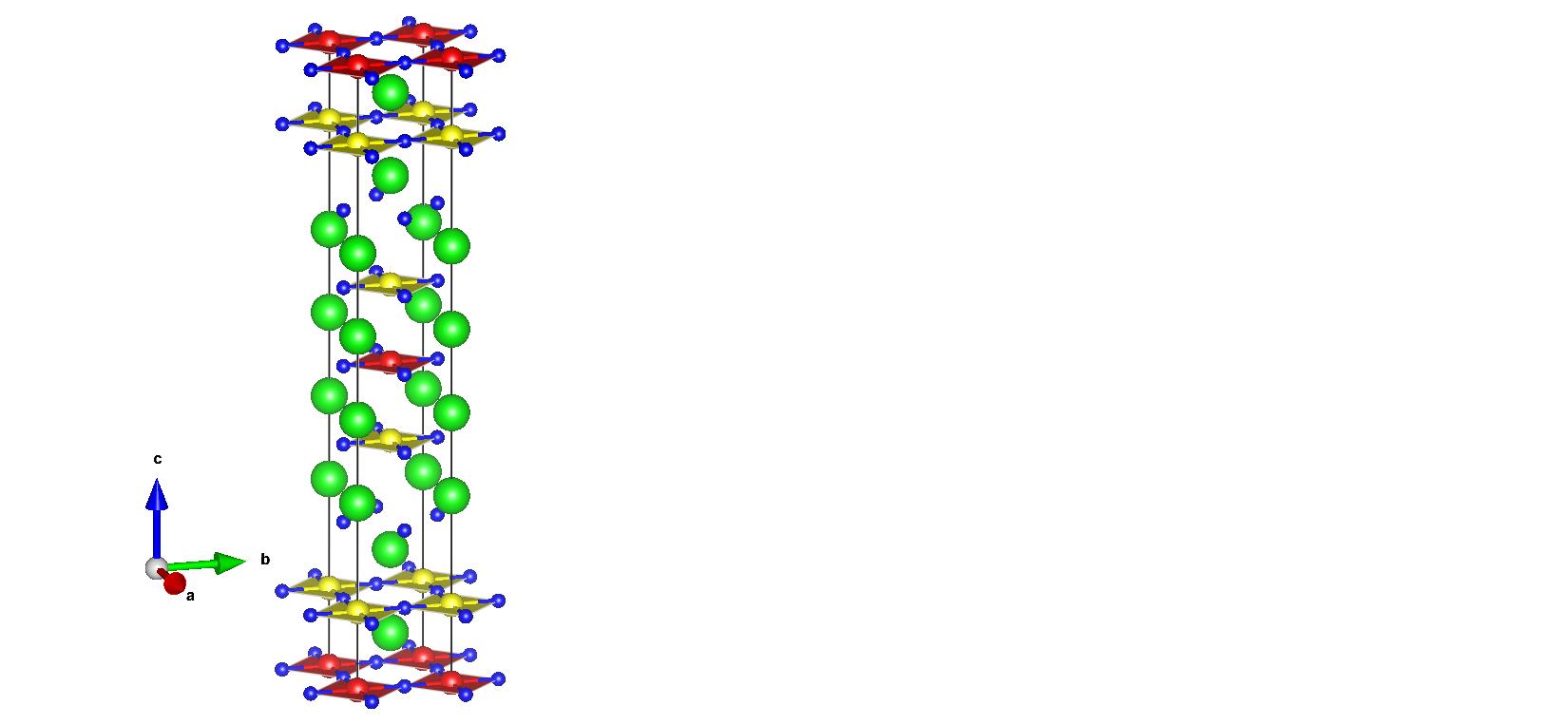}
\caption{\small(Color online) The crystal structure of La${_4}$Ni${_3}$O${_8}$. The NiO${_4}$ square planes are in red and yellow with Ni1 and Ni2 atoms in the centers, respectively. La atoms are represented as green spheres (big) and O atoms as blue spheres (small).}
\label{fig1}
\end{figure}

La${_4}$Ni${_3}$O${_8}$ crystallizes in a body centered tetragonal structure~\cite{Poltavets2007a} with space group I4/mmm (\#139). As appeared in Fig \ref{fig1}, this structure comprises of triple NiO$_2$ infinite layers separated by La${^{3+}}$ cations without oxygen, bringing planar nature to the crystal structure. These triple NiO$_2$ layers are separated from each other by La/O$_2$/La fluorite-type layers. The NiO$_2$ trilayer is made up of two types of Ni. The type$-$1 nickel (Ni1) and type$-$2 nickel (Ni2) are in inner and outer NiO$_2$ planes, respectively.  Using the experimentally determined structural parameters as input, structural optimization has been performed for La${_4}$Ni${_3}$O${_8}$ as well as for Sr substituted La${_4}$Ni${_3}$O${_8}$. The experimental lattice parameters are well reproduced by our GGA+\textit{U} calculations with only -0.5\% variation in equilibrium volume and 0.17\% variation in the Ni$-$Ni and Ni$-$O bond lengths in La${_4}$Ni${_3}$O${_8}$. The theoretically optimized structural parameters are compared with available experimental values in Table \ref{Table-I}.  The inner NiO$_2$ plane has an intralayer Ni1$-$O separation of 1.9817\AA\ and an O$-$Ni1$-$O angle of 180$^\circ$, whereas the external NiO$_2$ plane has a little distortion with an intralayer Ni2$-$O separation of 1.9819\,\AA\  bringing the O$-$Ni2$-$O angle to about 178$^\circ$. The presently calculated interlayer Ni1$-$Ni2 distance at the equilibrium volume is 3.2578\,\AA\ which is consistent with the previously reported value by Ting Liu \textit{et al.}~\cite{Liu2012}, obtained using similar structural optimization approach. The structural parameters for undoped and Sr doped La${_4}$Ni${_3}$O${_8}$ are given in Table \ref{Table-I}.
\section{Computational details}

\begin{figure*}[!t]
\centering
\captionsetup{justification=raggedright,singlelinecheck=on}
\begin{subfigure}{.24\linewidth}
    \centering
    \includegraphics[scale=0.2]{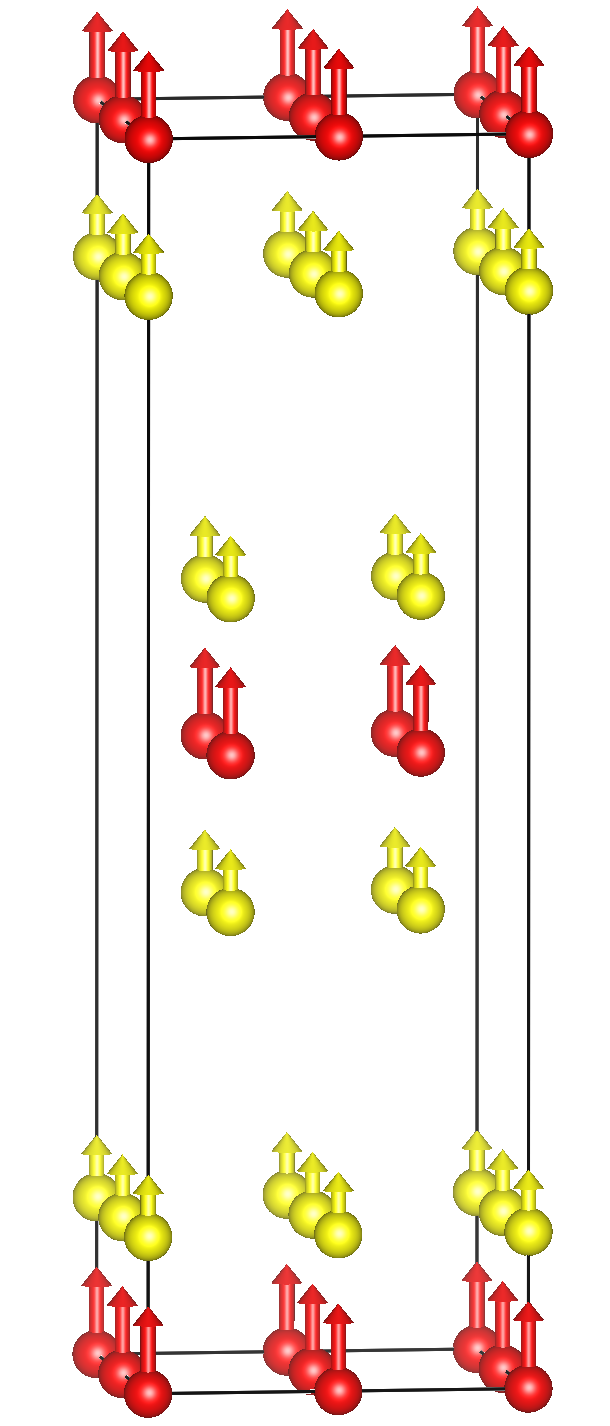}
    \caption{}
\end{subfigure}
\begin{subfigure}{.24\linewidth}
    \centering
    \includegraphics[scale=0.2]{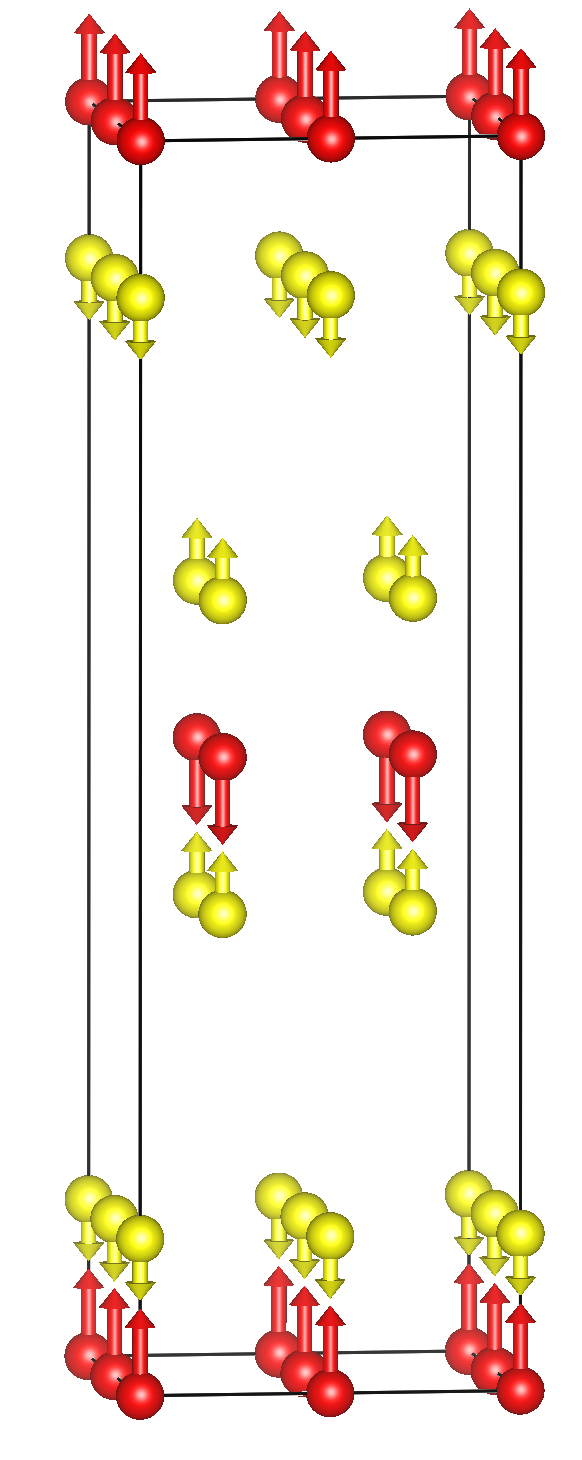}
    \caption{}
\end{subfigure}
\begin{subfigure}{.24\linewidth}
    \centering
    \includegraphics[scale=0.2]{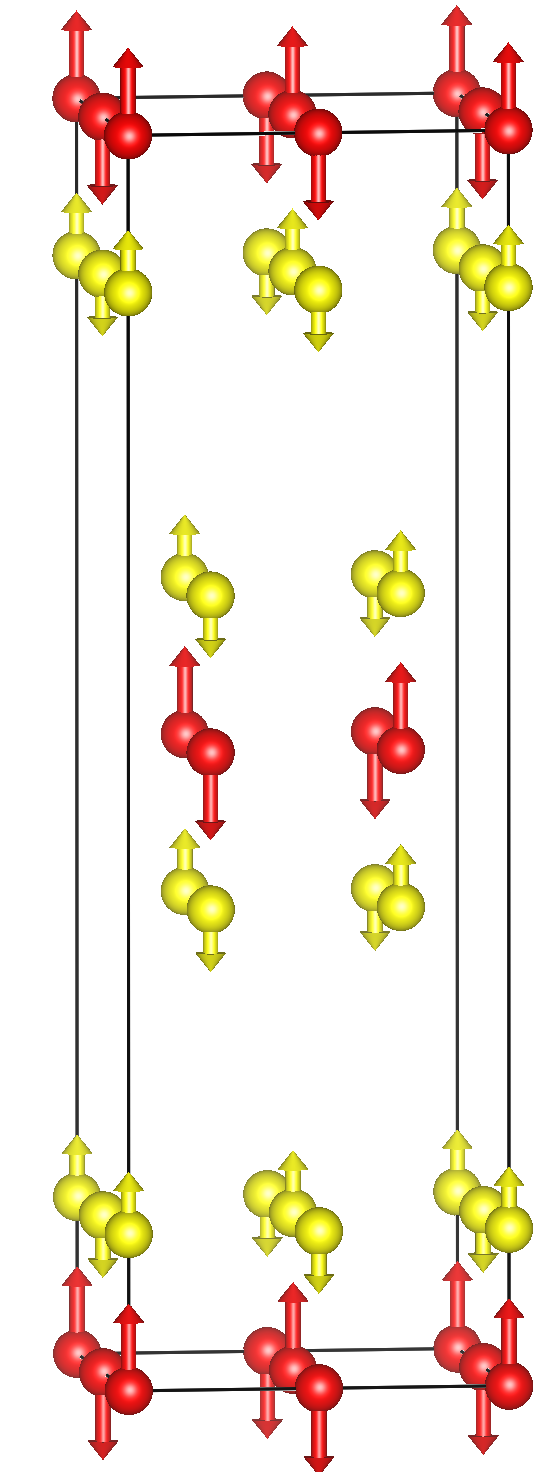}
    \caption{}
\end{subfigure}
\begin{subfigure}{.24\linewidth}
    \centering
    \includegraphics[scale=0.2]{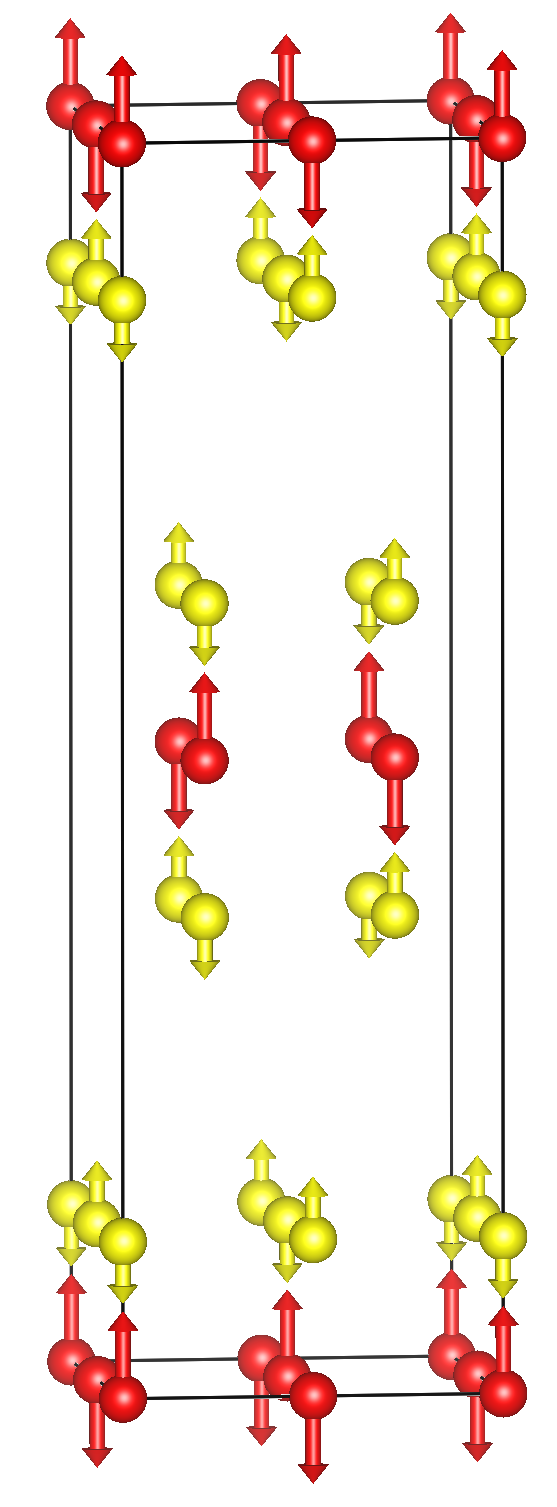}
    \caption{}
\end{subfigure}
\caption{\small(Color online) Schematic representation of (a) FM, (b) \textit{A}$-$AFM, (c) \textit{C}$-$AFM, and (d) \textit{G}$-$AFM ordering in a 2 x 2 x 1 supercell of tetragonal La${_4}$Ni${_3}$O${_8}$; only Ni ions are shown. Note that the supercell consists of two types of Ni atoms denoted as Ni1 (red spheres) and Ni2 (yellow spheres) where Ni1 (having large arrows) has larger magnetic moment than Ni2 (having small arrows).}
\label{fig2}
\end{figure*}

The calculations presented in this paper were performed using the standard full potential linearized augmented plane wave method implemented in the WIEN2k~\cite{Blaha2014, Schwarz2003}code based on density functional theory.~\cite{Jones1989} The room temperature powder neutron diffraction lattice parameters and the atomic positions were taken from Ref.~\cite{Poltavets2007}. For all calculations, the muffin$-$tin radii (R$_{mt}$) were chosen as 2.32, 2.35, 1.97 and 1.75\,a.u. for Sr, La, Ni and O, respectively. The cut$-$off parameter R$_{mt}$K$_{max}$ was set to 7.0. To ensure the convergence for the Brillouin zone integration, 1000 \textbf{k}$-$points were used over the irreducible wedge of the first Brillouin zone (IBZ). Self$-$consistency was accomplished by demanding the convergence of the total energy to be smaller than 10$^{-5}$ Ry/cell. This is comparable to a charge convergence to below 10$^{-4}$ electrons/atom. The generalized gradient approximation (GGA) using the parameterized scheme of Perdew \textit{et al.}~\cite{Perdew1996} was utilized for the exchange$-$correlation potential. The low dimesional transition metal oxides such as La${_4}$Ni${_3}$O${_8}$ will have strong Coulomb correlation effect arising from the transition metal 3$d$ states. In order to properly describe the strong electron correlation associated with Ni 3\textit{d} states, the GGA+\textit{U} method was used in the $'$fully localized limit$'$ (FLL).~\cite{Mazin2003} Here $U_{eff}$ = \textit{U$-$J} (\textit{U} and \textit{J} are on$-$site Coulomb repulsion and exchange interaction, respectively) was used instead of \textit{U}. For nickelates, the reasonable range of the \textit{U} parameter is 4$-$8\,eV.~\cite{Anisimov1999, Pardo2012, Poltavets2010, Pardo2010} So, the results presented below were calculated with $U_{eff}$ = 5\,eV and 6\,eV. To clarify the magnetic ground state, 2 x 2 x 1 supercell calculations were performed for one ferromagnetic (FM, shown in Fig. \ref{fig2}(a)) and three antiferromagnetic (AFM) spin configurations. For AFM configuration, at least three different magnetic arrangements are possible according to interplane and intraplane coupling, viz: (i) With interplane AFM coupling and intraplane FM coupling, the \textit{A}$-$AFM structure emerges as shown in Fig. \ref{fig2}(b). (ii) The inverse structure of \textit{A}$-$AFM, where the interplane coupling is FM and the intraplane coupling is AFM, is known as \textit{C}$-$AFM (Fig. \ref{fig2}(c)). (iii) If both the interplane and intraplane couplings are AFM, the \textit{G}$-$AFM~\cite{Ravindran2002} structure emerges as shown in Fig. \ref{fig2}(d).
\begin{table*}[!b]
\begin{threeparttable}
\centering
\caption{The structural parameters for the tetragonal  La${_{4-x}}$Sr$_x$Ni${_3}$O${_8}$ ($x$ stands for Sr substitution which varies from 0 to 3) are given where $a$ and $c$ represent the lattice parameters of the tetragonal structure (in \AA) and V represents the volume of the unit cell (in \AA$^3$).}
\label{Table-I}
\setlength{\tabcolsep}{14pt}
\setlength\extrarowheight{5pt}
\begin{tabular}{@{}lllll@{}}

\hline
  & $x$ = 0   & $x$ = 1   & $x$ = 2   & $x$ = 3   \\ \hline
V & 409.56  & 403.38  & 407.50  & 411.62  \\
&   411.12\tnote{a}                                \\
a & 3.9642  & 3.9441  & 3.9575  & 3.9708  \\
&   3.9708\tnote{a}                       \\
&   3.9705\tnote{b}                                 \\
c & 26.0621 & 25.9305 & 26.0184 & 26.1057 \\
& 26.1057\tnote{a}                      \\
& 26.103\tnote{b}                               \\ \hline
\end{tabular}
\begin{tablenotes}
            \item[a] Reference~\cite{Poltavets2007}
            \item[b] Reference~\cite{Blakely2011}
        \end{tablenotes}
\end{threeparttable}
\end{table*}
\section{Results and discussions}
\subsection{Electronic and magnetic structure of undoped La${_4}$Ni${_3}$O${_8}$}
The relative total energies of various spin configurations with respect to the ground state configuration are calculated and summarized in Table \ref{Table -II}. Our GGA calculations for all the considered spin configurations show metallic features which is in disagreement with the experimental observation.~\cite{Poltavets2010} However, within GGA, the FM and \textit{A}$-$AFM states are found to have lower energies than \textit{C}$-$AFM state by 35\,meV and 41\,meV, respectively, as given in Table \ref{Table -II}. The \textit{G}$-$AFM state is having very high energy than all the considered magnetic configurations and is about 250\,meV higher than the ground state i.e. \textit{A}$-$AFM state. It is well known that La${_4}$Ni${_3}$O${_8}$ is a strongly correlated material. Owing to the limitation of usual density-functional calculations in predicting the properties of strongly correlated materials, insulating transition metal oxides are generally predicted to be metals. This can be remedied by going beyond GGA such as GGA+\textit{U} calculations. Hence, in order to account for the strong correlation effect in La${_4}$Ni${_3}$O${_8}$, we have considered the correlation effect through including a Hubbard \textit{U} into the Hamiltonian matrix, and optimized the crystal structure for different magnetic structures using \textit{U} values of 6\,eV as well as 7\,eV.  Within GGA+\textit{U}, the calculation for \textit{G}$-$AFM state does not converge, consistent with observation made by Ting Liu \textit{et al.}~\cite{Liu2012} The calculated total energy and magnetic moments are listed in Table 1 except for the unconverged \textit{G}$-$AFM state. Within GGA+\textit{U} ($U_{eff}$ =5\,eV), the \textit{C}$-$AFM is the most stable configuration. The energy of \textit{C}$-$AFM is about 190\,meV and 180\,meV lower than the FM and \textit{A}$-$AFM states, respectively. Also only the \textit{C}$-$AFM configuration shows insulating behavior, which is consistent with the experimental findings. As covalent bonding is present between O 2\textit{p} and Ni 3\textit{d} orbitals, one could expect induced magnetic moment in the oxygen site. Our GGA+\textit{U} calculation shows that, in the FM configuration for the undoped system, the magnetic moment present in the neighboring oxygen sites are 0.040\,$\mu_B$, 0.039\,$\mu_B$ and 0.002\,$\mu_B$ respectively, for O1, O2 and O3 atoms. As O1 and O2 are the oxygens present within the square planar layer, they have a strong hybridization with Ni1 and Ni2, respectively. But O3 atom is present in the fluorite-type layer which has no direct interaction with Ni atoms resulting in lower magnetic moment than the other two oxygens. The magnetic moments at the oxygen ions are found to be directed parallel to the minority$-$spin channel of Ni atoms.
\par
One can usually expect large magnetic moment in the FM configuration, compared to the AFM configuration. In the present system we have found that the magnetic moment in the Ni site for the FM configuration is comparable with that of \textit{A}$-$AFM configuration. So the magnetic moment in the Ni site is influenced by the intralayer FM coupling between the Ni atoms. It is expected that, due to the layered structure of this system, the interlayer FM coupling will be weaker and hence it may not influence the ground state magnetic configuration. Our calculation shows that the interlayer FM coupling is very important to understand the stability of this system. In fact, the calculated magnetic moments in the Ni sites are very small in the case of \textit{C}$-$AFM if the strong correlation effect is not included into the calculation through the \textit{U} parameter. However, when we include strong Coulomb correlation into the calculation using GGA+\textit{U} method, the magnetic moment at the Ni sites increases by several order of magnitudes in the \textit{C}$-$AFM case. But the magnetic moment enhancement in the FM and \textit{A}$-$AFM configuration is not that large as compared to the enhancement in \textit{C}$-$AFM configuration. Most importantly, the experimentally found insulating behavior could be reproduced only when we include the strong correlation effect into the calculation. This clearly indicates that the origin of the insulating behavior in La${_4}$Ni${_3}$O${_8}$ is not only due to the FM coupling between the layers but also due to the Coulomb correlation effect present in the system.

\begin{table*}[]
 \begin{threeparttable}
\centering
\caption{The calculated total energies $\triangle$E (meV/f.u.) relative to the lowest energy states, the magnetic moments M$_s$ ($\mu_B$) at the Ni sites, band gap E$_g$ (eV) values for different magnetic states depend on the \textit{U} (eV) values for La${_4}$Ni${_3}$O${_8}$. }
\label{Table -II}
\setlength{\tabcolsep}{18pt}
\setlength\extrarowheight{5pt}
\begin{tabular*}{\textwidth}{@{\extracolsep{\fill}}llllll}
\hline
\textit{U}                    &                      & FM         & \textit{A}$-$AFM      & \textit{C}$-$AFM      & \textit{G}$-$AFM                \\ \midrule
0                    & $\triangle$E                   & 5.9        & 0          & 40.95      & 253.83               \\
                     & M$_s$                   & 0.52/0.54  & 0.50/0.52  & 0.27/0.29  & 0.21/0.28            \\
\multicolumn{1}{l}{} &  & 0.52/0.54\tnote{a} & 0.57/0.52\tnote{a} & 0.29/0.27\tnote{a} &                      \\
                     & E$_g$                   & metal      & metal      & metal      & metal                \\
5                    & $\triangle$E                   & 189.28     & 179.46     & 0          &                      \\
                     & M$_s$                   & 0.77/0.75  & 0.81/0.78  & 1.44/1.29  &                      \\
\multicolumn{1}{l}{} &  & 0.79/0.77\tnote{a} & 0.83/0.81\tnote{a} & 1.44/1.29\tnote{a} &                      \\
\multicolumn{1}{l}{} &  & 0.79/0.77\tnote{b} & 0.8/0.8\tnote{b}   & 1.44/1.29\tnote{b} &                      \\
                     & E$_g$                   & metal      & metal      & 0.7        &                      \\
\multicolumn{1}{l}{} & &            &            & 0.7\tnote{b}       &                      \\
6                    & $\triangle$E                   & 340.65     & 330.73     & 0          &                      \\
                     & M$_s$                   & 0.80/0.79  & 0.83/0.80  & 1.50/1.34  &  \\
\multicolumn{1}{l}{} &  & 0.82/0.79\tnote{a} & 0.85/0.80\tnote{a} & 1.49/1.35\tnote{a} &                      \\
\multicolumn{1}{l}{} &  &            &            & 1.39/1.25\tnote{c} &                      \\
                     & E$_g$                   & metal      & metal      & 0.58       &  \\ \hline
\end{tabular*}
\begin{tablenotes}
            \item[a] Reference~\cite{Liu2012}
            \item[b] Reference~\cite{Liu2014}
            \item[c] Reference~\cite{Wu2013}
        \end{tablenotes}
\end{threeparttable}
\end{table*}

\par
In order to understand the role of various magnetic orderings on electrical properties of  La${_4}$Ni${_3}$O${_8}$ we have analyzed in detail the total and orbital projected density of states (DOS) for this system in different magnetic configurations. The optimized structure for La${_4}$Ni${_3}$O${_8}$ shows that both Ni1 and Ni2 are surrounded by four oxygen atoms in square-planar co-ordination. In an ideal square-planar crystal field, Ni 3\textit{d} splits into the highest level $x^2-y^2$, middle levels consist of doubly degenerate $z^2$, \textit{xy} and the lowest levels consist of doubly degenerate \textit{yz}, \textit{xz}.~\cite{Wells1962} But, due to the layered structure of La${_4}$Ni${_3}$O${_8}$, the energy of $z^2$ orbital is lowered. This is because of the reduced Coulomb repulsion for both Ni1 and Ni2 in the layers.~\cite{Xiang2008} In such a picture, d$_{x^2-y^2}$ orbital of the Ni atom will form sigma bond with O-2$p$ orbitals. Due to the absence of apical oxygens d$_{z^2}$ orbital forms weak pi bond with O-2$p$ orbitals. As the \textit{xy}, \textit{yz} and \textit{xz} orbitals are fully occupied, they hybridize weakly with O-2$p$ orbitals of oxygen atom (Fig.\ref{fig4}). The orbital projected DOS analysis shows that the hybridization between Ni-3$d$ and O-2$p$ orbitals is important to explain the electronic structure of La${_4}$Ni${_3}$O${_8}$. Our site- and angular-momentum-projected DOS for the ground state \textit{C}$-$AFM configuration shows that the DOS for Ni-3$d$ and O-2$p$ orbitals are degenerate in the energy range from $-$6 eV to E$_F$ which ensures covalent bonding between these two atoms. Because of this strong covalent hybridization of Ni 3\textit{d} with neighboring O 2\textit{p} states, the calculated DOS curves shown in fig. 5 show broad features. In order to understand the effect of square planar geometry on Ni 3\textit{d} orbitals, the orbital projected DOS is plotted which is shown in Fig. \ref{fig4}.
\begin{figure}[!b]
\includegraphics[width=8.5cm, height=9cm]{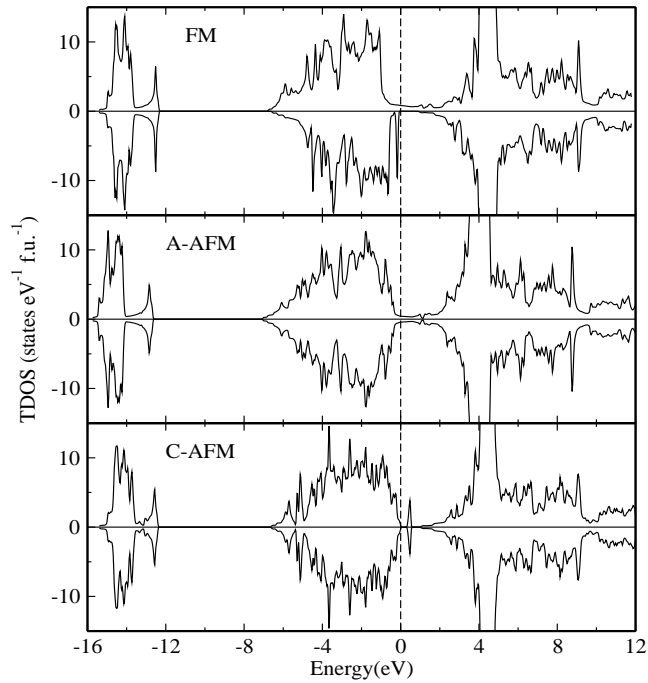}
\setlength{\belowcaptionskip}{0.5 pt}
\caption{\small Total DOS for La${_4}$Ni${_3}$O${_8}$ in ferromagnetic (FM), \textit{A}$-$type antiferromagnetic (\textit{A}$-$AFM) and \textit{C}$-$type antiferromagnetic (\textit{C}$-$AFM) states obtained from GGA+\textit{U} calculations with $U_{eff}$ = 5\,eV.}
\label{fig3}
\end{figure}
\begin{figure}[!t]
\setlength{\belowcaptionskip}{-1 pt}
\includegraphics[width=8.5cm, height=10cm]{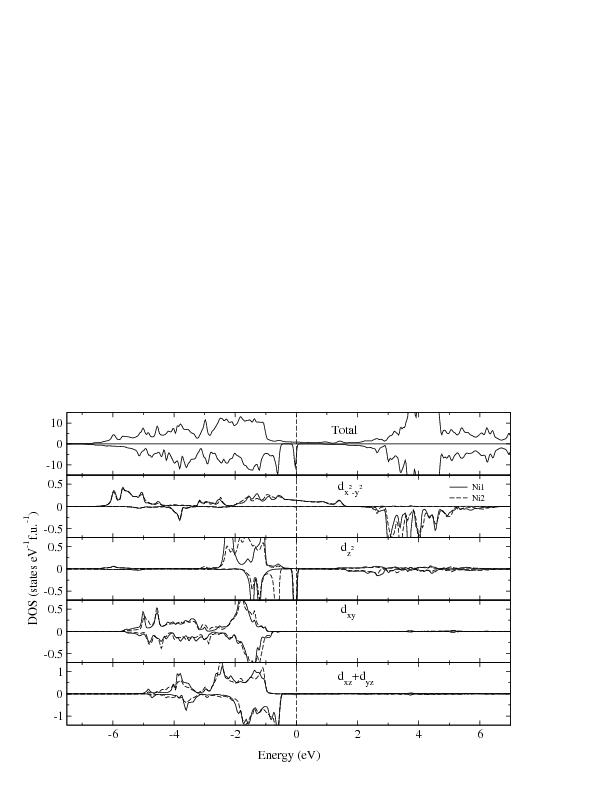}
\caption{\small Orbital projected DOS of La${_4}$Ni${_3}$O${_8}$ in FM configuration calculated using GGA+\textit{U} method with $U_{eff}$ = 5\,eV}
\label{fig4}
\end{figure}

\par
\begin{figure}[!t]
\includegraphics[width=8.7cm, height=17cm]{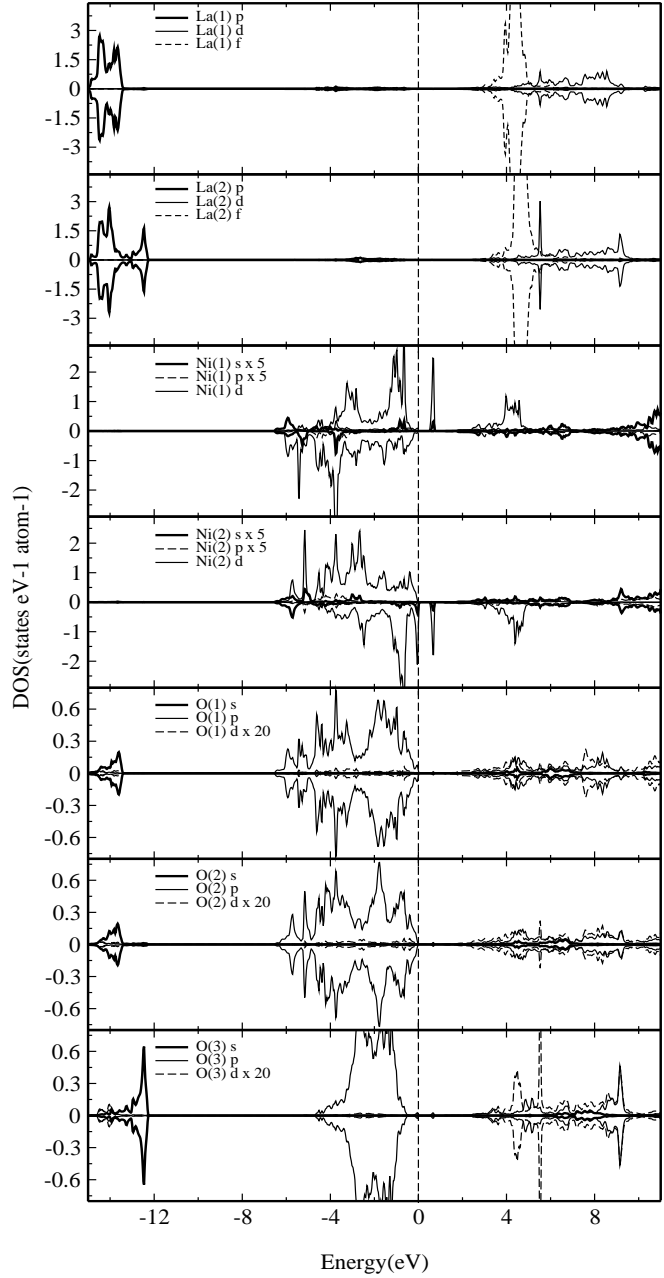}
\setlength{\belowcaptionskip}{0.5 pt}
\caption{\small Site and angular-momentum projected DOS for La${_4}$Ni${_3}$O${_8}$ \textit{C}$-$AFM configuration calculated using GGA+\textit{U} method with $U_{eff}$ = 5\,eV}
\label{fig5}
\end{figure}

Let us first analyze the results from the calculations corresponding to the FM configuration. The orbital projected DOS obtained from the GGA+\textit{U} ($U_{eff}$ = 5\,eV) is shown in Fig.\ref{fig4} for the FM configuration. La${_4}$Ni${_3}$O${_8}$ exhibits half metallicity in the FM state as seen in the Fig. \ref{fig3} and Fig. \ref{fig4} (with metallic character in up$-$spin channel and semiconducting character with a band-gap of 0.86\,eV in the down$-$spin channel). The majority spin  $x^2-y^2$ states with broad band features and significant bands from O 2\textit{p} states cross the Fermi level (not shown) resulting in half-metallicity. As seen from the figure~\ref{fig4}, the 3\textit{d} DOS distribution for the Ni1 and Ni2 have almost the same shape, and also have the same electron occupation. This implies that Ni atoms have a charge homogeneous solution with the average valence of Ni$^{+1.33}$ in this system. Consistent with this view point, the calculated magnetic moment in both the Ni sites are almost the same. It can be noted that the presence of half metallic behavior in the FM configuration disagrees with the experimental observation of insulating behavior. From the figure 4, we found that the majority$-$spin $x^2-y^2$ orbitals of Ni1 and Ni2 are partially (2/3) occupied, while the minority$-$spin $x^2-y^2$ orbitals of both the ions are nearly empty. The other 3\textit{d} orbitals are completely filled which suggests that both the ions hold the low spin state in the FM configuration. Each 2/3 occupied up$-$spin $x^2-y^2$ band of Ni1 and Ni2 ions contribute 2/3\,$\mu_B$ to the total calculated moment of 2\,$\mu_B$ per formula unit. As the $d_{x^2-y^2}$ orbital has an in$-$plane orientation, an \textit{A}$-$type antiferromagnetic state (intra$-$layer FM and inter$-$layer AFM within the trilayer) turns out to have a similar DOS characteristic (not shown here) as that of FM configuration mentioned above. So the \textit{A}$-$AFM configuration also possesses a low spin metallic solution.
From the calculated total energy for La${_4}$Ni${_3}$O${_8}$ in various magnetic configurations given in Table \ref{Table -II} it is clear that  the \textit{C}$-$AFM state is the most stable state within GGA+\textit{U}. As the Fermi level falls in the gap for both spin channels, the \textit{C}$-$AFM configuration exhibits an insulating feature (0.7\,eV with $U_{eff}$= 5\,eV). The valence band DOS of La${_4}$Ni${_3}$O${_8}$ are primarily derived from Ni 3\textit{d} and O 2\textit{p} admixture. The negligibly small DOS contribution in the valence band (VB) at the La site indicates pronounced ionic bonding between La and other constituents. The almost same topology of the DOS profile for the two Ni sites strongly suggests average valence states for both Ni atoms. In Fig. \ref{fig5} we have displayed the angular momentum projected DOS for La, Ni and O atoms in La${_4}$Ni${_3}$O${_8}$. The \textit{s} and \textit{p} states of Ni  and \textit{d} states of O  have negligible contribution (visible only after significant magnification) in the valence band in the vicinity of E${_F}$.
\par
\begin{figure}[!t]
\includegraphics[width=8.7cm, height=9cm]{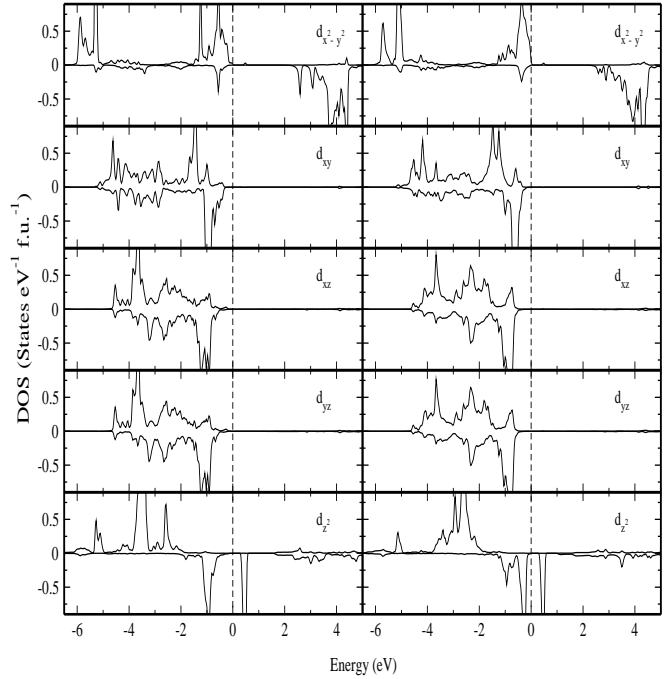}
\setlength{\belowcaptionskip}{0.5 pt}
\caption{\small Orbital projected DOS for Ni (Ni1 left and Ni2 right) in \textit{C}$-$AFM  state for La${_4}$Ni${_3}$O${_8}$ calculated by GGA+\textit{U} method with $U_{eff}$ = 5\,eV}
\label{fig6}
\end{figure}

The \textit{d} states of Ni atoms and \textit{p} states of O1, O2 atoms distributed from $-$6\,eV to E${_F}$. This implies that the Ni \textit{d} and O \textit{p} states are strongly hybridized. The O 2\textit{s} states for O1 and O2 are well localized and are present around $-$15\,eV. As O3 is present in the La2/(O3)$_2$/La2 fluorite type layers, it has somewhat different DOS characteristics as compared to the square planar oxygens (O1 and O2). The 2\textit{s} states for O3 atom are present in a higher state as compared to that for the other two oxygen atoms. The \textit{xy}, \textit{yz} and \textit{xz} orbitals of Ni atoms are fully occupied and localized in a narrow energy range indicating that they hybridize weakly with the O 2\textit{p} orbitals. Because of the absence of apical oxygens, the $d_{z^2}$ orbital forms pd$\pi$ bonding with O 2\textit{p} orbital. The $d_{x^2-y^2}$ orbital hybridizes strongly with the O 2\textit{p} orbital within the layers to produce broad bands. The Ni 3\textit{d} states produce narrow bands which indicate the strong correlation effect in La${_4}$Ni${_3}$O${_8}$. As shown in Fig. \ref{fig6}, the majority spin DOS for $d_{z^2}$ orbital is completely occupied, whereas the DOS for the corresponding orbital in the minority spin state is partially occupied. It may be noted that the occupied and unoccupied levels are separated by a gap of around 0.7\,eV. However it is interesting to note that our GGA+\textit{U} calculations give rise to some intermediate band gap states. The present result is, to the best of our knowledge, the first to point out the occurrence of intermediate energy states in the band-gap of  La${_4}$Ni${_3}$O${_8}$. Intermediate band-gap oxides attract much attention due to their potential applications in solar cells.~\cite{PhysRevLett.78.5014, 10.1039/9781849739955-00425} The orbital projected DOS shows that these intermediate states are originating from $d_{z^2}$ electrons from both the Ni ions which may be induced by the correlation effect.

\par
Let us now turn our attention to the valence and spin states of Ni ions in La${_4}$Ni${_3}$O${_8}$ in the ground state magnetic configuration. Due to the Coulomb repulsion induced by four $d_{z^2}$ electrons of two Ni2 ions within the trilayer along the c$-$axis, one Ni1 $d_{z^2}$ electron will be excited to the corresponding $d_{x^2-y^2}$ orbital. As a result all the nickel ions would have half-filled $d_{x^2-y^2}$ orbital and ultimately \textit{C}$-$AFM would be in a high spin state. So both the $d_{x^2-y^2}$ and $d_{z^2}$ orbitals should have one hole in the minority spin state which can be seen from the orbital projected DOS given in Fig. \ref{fig6}. From the integrated orbital projected DOS, we have calculated the occupation numbers for the majority spin of $d_{x^2-y^2}$ and $d_{z^2}$ orbitals of Ni1 ion which are 0.95$e$ and 0.92$e$, respectively, as expected. But the strong hybridization between Ni 3\textit{d} and O 2\textit{p} orbitals induces 0.17$e$ and 0.36$e$ for the minority spins states of the $d_{x^2-y^2}$ and $d_{z^2}$ orbitals, respectively. All other \textit{d}-orbitals of Ni1 ions are found to be fully occupied. Similarly, for the two Ni2 ions present in the trilayer, the $d_{x^2-y^2}$ majority (minority) spin state and the $d_{z^2}$ majority (minority) spin state have the occupations numbers of 0.92$e$ (0.16$e$) and 0.89$e$(0.45$e$), respectively. As a consequence of this, the minority spin $d_{z^2}$ electron of Ni2$^{1+}$ can readily hop to the formally empty minority spin $d_{z^2}$ orbital of Ni1$^{2+}$ ion. This stabilizes the interlayer FM coupling within the trilayer. The band originating from $x^2-y^2$ states in the \textit{C}$-$AFM configuration is narrower  compared to that in the FM configuration. This reduction in band width is due to the weaker intralayer AFM coupling. In \textit{C}$-$AFM configuration, the Ni atoms align antiferromagnetically due to linear Ni$-$O$-$Ni superexchange interactions.~\cite{Goodenough1963} The $d_{x^2-y^2}$ orbital of Ni atoms interact through the same oxygen $p_x$/$p_y$ orbital causing antiparallel spins on neighboring Ni atoms and this could explain the stabilization of the AFM configuration over the FM configuration.
\par
The next puzzle is the understanding the origin of comparatively larger moment at the Ni1 site than that in the Ni2 site. If an oxygen atom is removed from a Ni$-$O$-$Ni configuration with bond angle 180$^\circ$, the Ni atoms repel each other due to the electrostatic force and the neighboring O atoms collectively move towards the vacancy to screen the repulsion. The optimized crystal structure parameters show that the Ni atoms of the Ni1 site are surrounded by four base-plane O atoms at a distance of 1.9817\AA\ and the corresponding distance is 1.9819\AA\ in case of Ni2 site. These findings strongly suggest that the bonding interaction between Ni and O is stronger in the Ni1O$_4$ coordination than that in Ni2O$_4$ coordination. Hence, compared with the local surrounding of Ni2O$_4$, relatively more of the electrons in the Ni1O$_4$ subunits are participating in bonding rather than in magnetism. Since the bond lengths are almost similar for Ni1$-$O and Ni2$-$O, the magnetic moments at Ni1 and Ni2 are expected to be the same. But our calculation shows that the magnetic moment at the Ni1 site is 1.44\,$\mu_B$ which is comparatively larger than that at the Ni2 site (1.29\,$\mu_B$). This is because of the small distortion in the Ni2O$_4$ site for which the Ni$-$O$-$Ni bond angle is 178$^\circ$ vs 180$^\circ$ in the case of Ni1O$_4$. Due to the reduction in bond angle, the Ni2$-$O$-$Ni2 super-exchange is hindered resulting in a lower magnetic moment at the Ni2 site.

\subsection{Role of hole doping La$_{4}$Ni$_{3}$O$_{8}$}
In order to understand the changes in the electronic structure and magnetic properties of La$_{4}$Ni$_{3}$O$_{8}$ due to hole$-$doping, we have considered divalent Sr substitution (hole doping) at the La site. We  have carried out calculations for this hole doped systems in different magnetic configurations similar to the undoped case mentioned above, adopting the similar approach. As there are several possibilities to replace La with Sr in La$_{4}$Ni$_{3}$O$_{8}$ we have substituted La by Sr atom through supercell approach such that the Sr will locate with respect to the plane in the middle of the square spanned by the four Ni ions in the supercell.
\begin{figure}[!t]
\includegraphics[width=8.5cm, height=10cm]{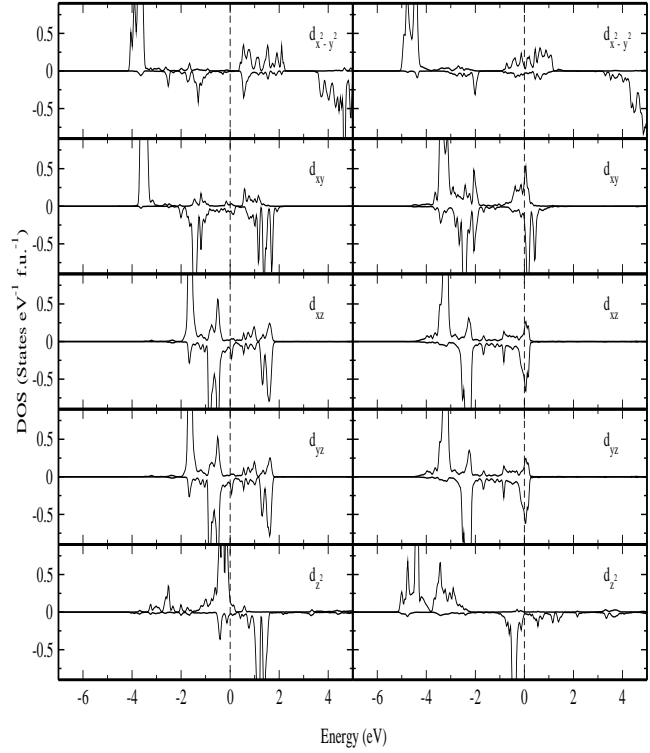}
\caption{\small Orbital projected DOS for Ni (Ni1 left and Ni2 right) close to Sr in \textit{C}$-$AFM state for La${_3}$SrNi${_3}$O${_8}$ calculated by GGA+\textit{U} method with $U_{eff}$ = 5\,eV}
\label{fig7}
\end{figure}
\subsubsection{Structural changes}

 No systematic composition dependence of the parameters was observed which can be seen from Table~\ref{Table-I}. But the increase in the lattice parameters from $x$ = 1 to $x$ = 3 can be understood as follows. The ionic radius of Sr$^{2+}$ is slightly larger than that of La${^{3+}}$. Hence, when La${^{3+}}$ is replaced by Sr${^{2+}}$ a negative chemical pressure acts primarily on the NiO$_2$ planar subunits which increases the Ni$-$O$-$Ni bond angles. In order to compensate for the local repulsion between overlapping charge densities, the Ni$-$O bond lengths increase. In the case of La$_{4-x}$Sr$_x$Ni$_{3}$O$_{8}$, for $x$ = 1 and 2, Sr${^{2+}}$ occupies the vacancy created by the removal of La${^{3+}}$. And in the case of $x$= 3, when the La/O$_2$/La layer is replaced by Sr, a further increase in  the cell volume is noticed (Table 1). The observed c/a ratio has a maximum value at $x$ = 2, because holes are first doped into the $d_{x^{2}-y^{2}}$ orbital for lower concentration and, then, additional holes are doped into the $d_{z^2}$ orbital for higher concentration of Sr similar to the bilayer system investigated by  Gopalkrishnan \textit{et al.}~\cite{Gopalkrishnan1977}  and Takeda \textit{et al.}.~\cite{Takeda1990} Moreover, since Sr doping is equivalent to hole doping, the number of electrons contained in VB decreases with increasing  Sr concentration. This leads to a systematic shift of E${_F}$ towards the lower energy region of the DOS.
As size of Sr atom is not very different from La (ionic radii of Sr and La are 0.115 and 0.112 a.u. respectively), the hole doping does not modify the structure of the parent material significantly which can be seen from the optimized structural parameters given in Table 1. Because of the smaller electronegativity difference between Sr and La, the substitution of La with Sr will not have noticeable changes in the band structure except band filling originating from the hole-doping effect.

\begin{figure}[!t]
\includegraphics[width=8.5cm, height=17cm]{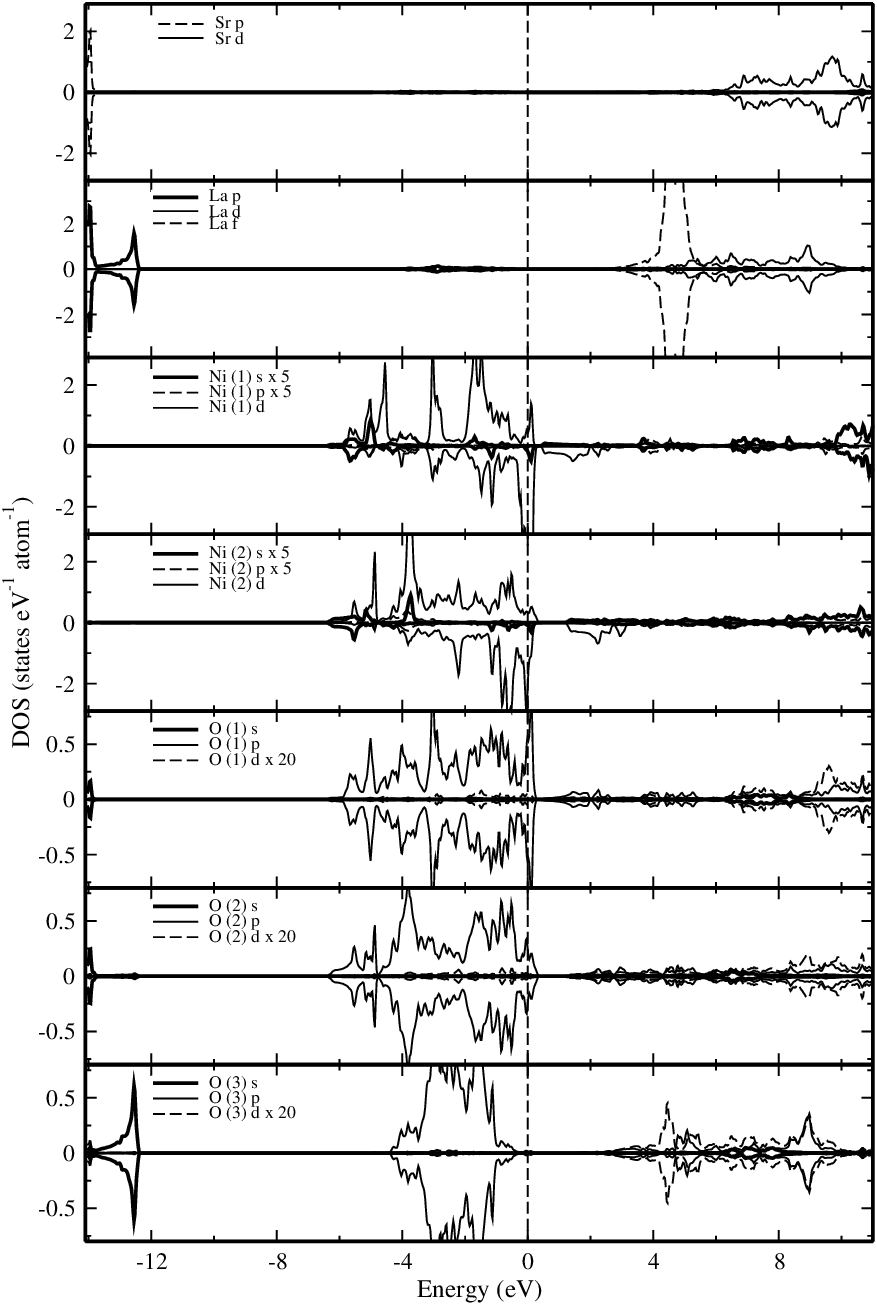}
\caption{\small Partial DOS for \textit{C}$-$AFM state of La${_2}$Sr${_2}$Ni${_3}$O${_8}$ calculated by GGA+\textit{U} method with $U_{eff}$ = 5\,eV}
\label{fig8}
\end{figure}

\subsubsection{ Changes in electronic and magnetic structure by hole-doping}

\begin{figure}[!t]
\includegraphics[width=8.5cm, height=10cm]{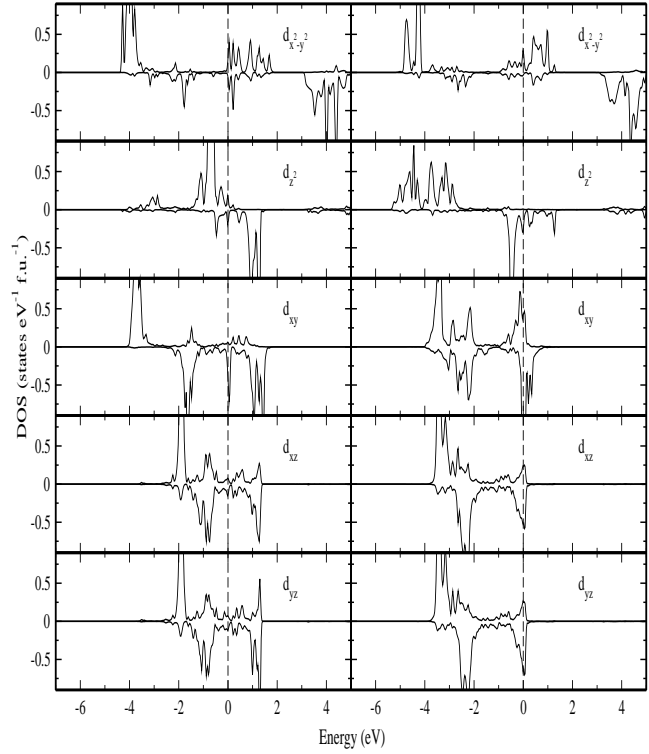}
\caption{\small Orbital projected DOS for Ni (Ni1 left and Ni2 right) close to Sr in \textit{C}$-$AFM state for LaSr${_3}$Ni${_3}$O${_8}$  calculated by GGA+\textit{U} method with $U_{eff}$ = 5\,eV}
\label{fig9}
\end{figure}

As Sr is divalent and La trivalent, removal of electrons from the NiO$_4$ sub-lattice is necessary to balance the charge deficiency introduced by the Sr substitution. Thus VB shows an upward shift towards the Fermi energy as shown in Fig. \ref{fig8}. From this figure it is clear that the small concentration of La substituted by Sr in La${_4 }$Ni${_3}$O${_8}$ significantly changes the electronic structure of the system.  Our calculated DOS curves also show significant changes in the electronic structure of both types of Ni atoms due to the substitution of La by Sr as evident in Fig. \ref{fig7} and Fig. \ref{fig9}. This indicates that hybridization and charge transfer effects play important roles in describing the electronic structure of La${_{4-x}}$Sr${_x}$Ni${_3}$O${_8}$. As the stability of La${_{4-x}}$Sr${_x}$Ni${_3}$O${_8}$ mainly originates from the covalent interaction between the Ni 3\textit{d} and O 2\textit{p} electrons, changing the $d$ band filling of the Ni $d$ states in the valence band around the Fermi level at higher Sr concentration is expected to give rise to structural destabilization. This could explain why, high Sr concentrations in La${_{2-x}}$Sr${_x}$NiO${_4}$ could always bring oxygen vacancy as reported experimentally~\cite{K1994}. Hence it is expected that the higher Sr concentration in La${_4 }$Ni${_3}$O${_8}$ also stabilize this system with oxygen vacancy.

\par
The calculated DOS for La${_2}$Sr${_2}$Ni${_3}$O${_8}$ (with 50\% Sr substitution) for the ground state \textit{C}$-$AFM configuration is given in Fig.\ref{fig8}. From the DOS analysis we have found that, except the localized states at around -12\,eV to -14\,eV  derived from Sr, the DOS in the whole energy range is almost the same as that of La${_4}$Ni${_3}$O${_8}$ irrespective of the Sr concentration. However, all the electronic states in the VB get shifted towards the higher energy upon Sr substitution as evident from the calculated DOS. This can be explained as follows. Since Sr doping is equivalent to removing one electron from the system, the number of electrons contained in the VB decreases with increasing amount of Sr concentration. This leads to a systematic shift of E${_F}$ towards the lower energy region of DOS, if the system follows simple band filling effect. It may be noted that, similar to La, Sr also donates all its valence electrons to the host lattice and forms ionic bonding. Replacing La with Sr introduces hole states above the Fermi level and reduces the electron per atom ratio in La${_4}$Ni${_3}$O${_8}$. If we assume that the rigid$-$band$-$filling principle works in this case, the reduction in the electron per atom ratio is equivalent to shifting E${_F}$ towards the lower energy side of DOS. As there are not very much changes in the DOS topology with Sr substitution, one can conclude that Sr substitution will only change the valence electron count without influencing the hybridization interaction between Ni and O significantly. So we focus on the position of the  Fermi level, width of the valence band as well as conduction band and  bandgap changes as a function of Sr substitution, because they play a key role in the electrical, magnetic, and optical property of the system.
\par
In order to understand the role of Sr substitution on electrical and magnetic properties of La${_4}$Ni${_3}$O${_8}$ we have plotted the orbital projected DOS for the constituents with  25\% Sr substitution (See Fig~\ref{fig7}). The orbital projected DOS shows that more of the \textit{d}$-$dos of the Ni1 site has moved towards the unoccupied site than the that in Ni2 site. This is not surprising, because two Sr atoms are present as neighbors for Ni1 whereas the Ni2 is surrounded by only one Sr ion alone in  La${_3}$SrNi${_3}$O${_8}$. In this system, it is natural that Sr substitution adds one hole in the Ni sites resulting in increase of total oxidation states of both Ni ions from  4+ to 5+ state. As the radius of Sr${^{2+}}$ is (slightly) larger than that of La${^{3+}}$, it is suggested that the Sr doping favors the HS state in Nickelates  by the introduction of partial oxidation due to covalency effect.~\cite{Munakata1997, Ravindran2002} As the doped system also possesses average valency (all Ni atoms are having almost equal moments), the partial oxidation state introduced by hole doping increases the average valency from +1.33 to +1.67 i.e. Ni atoms will have HS with  maximum magnetic moment of 1.67\,$\mu_B$. But these considerations correspond to the purely ionic model, the hybridization of Ni 3\textit{d} orbitals with O 2\textit{p} orbitals and the resulting broad band formation  can significantly renormalize this ionic value. Hence the calculated magnetic moments are always smaller than the ideal ionic values mentioned above. It may be noted that we have not considered O non-stoichiometry in the present calculations. But in experiment one may expect oxygen deficiency in hole-doped systems or at high temperatures. Especially  one can expect a disagreement between experiment and theory for $x$ > 0.5, where the oxygen deficiency is known to increase significantly. From the Fig. \ref{fig8}, we can see that, the reduction of electron population caused by hole-doping affects the electron distribution not only in  Ni1 and Ni2, but also in O1 and O2 atoms. Due to the hole-doping, the Ni and O DOS cross the Fermi level bringing insulator-to-metal transition without affecting the AFM ordering.
\par
According to Goodenough~\cite{Mcmorrow1976}, a change between FM and AFM states may be expected when the magnetic moment for the FM and AFM states is almost the same ($\mu_F\sim \mu_{AF}$); the stable magnetic phase being the one that has the highest magnetic moment. Our calculated results are consistent with this viewpoint in the sense that the magnetic moments are found to be larger for the stable magnetic configuration ($C$-AFM) in La${_{4-x}}$Sr${_x}$Ni${_3}$O${_8}$ (see Table~\ref{Table -II} for magnetic moment of different magnetic configurations for ${x}$= 0). When we go from La${_3}$SrNi${_3}$O${_8}$ to LaSr${_3}$Ni${_3}$O${_8}$, the magnetic moment increases. The increase in magnetic moment as a function of hole doping originates from two reasons: First, owing to band-filling effects the hole-doping moves the Fermi level to the lower energy side and thus enhances DOS at the Fermi level as well as the exchange splitting. Second, owing to the larger size of Sr${^{2+}}$ than La${^{3+}}$, lattice is expanded for increased Sr substitution leading to reduced hybridization and increased band localization. As a result, the bands become narrow, which in turn enhances the spin polarization. Furthermore, the increased exchange splitting of the Ni 3\textit{d} bands also enhances magnetic moment.
\par
In order to understand the change in the magnetic properties of the Ni ions present in the vicinity of Sr or La atoms, spin polarized Ni 3\textit{d} DOS for Ni1 and Ni2 atoms closer to Sr and La are presented in Fig. \ref{fig7} and Fig. \ref{fig9} for the La${_3}$SrNi${_3}$O${_8}$ and LaSr${_3}$Ni${_3}$O${_8}$ cases, respectively. From this illustration it is clear that the $d$ states for  both Ni atoms are itinerant and also that both are participating in the magnetism. It is interesting to note that Ni atoms closer to Sr are more spin polarized than those closer to La owing to the reasons discussed above. For the \textit{C}$-$AFM ground state of La${_3}$SrNi${_3}$O${_8}$, Ni1 closer to La possesses a magnetic moment of 1.50\,$\mu_B$ and that closer to Sr has 1.62\,$\mu_B$. Similarly, Ni2 closer to La possesses a magnetic moment of 1.34\,$\mu_B$ and those closer to Sr have 1.67\,$\mu_B$. In case of $x$=3 i.e. LaSr${_3}$Ni${_3}$O${_8}$, the combined oxidation state of Ni atoms (Ni1 + 2Ni2) is increased from 4+ to 7+. So we can say that the Ni1 is in 3+ (\textit{d$^7$}) and Ni2 is in 2+ (\textit{d$^8$}) oxidation states. But magnetic moments obtained in none of the Ni site reflect these oxidation states (e.g. Ni$^{2+}$ in HS is expected to be 1$\mu_B$). Therefore, signature for charge ordering with \textit{d$^7$}+2\textit{d$^8$} ions is not observed in LaSr${_3}$Ni${_3}$O${_8}$. The DOS distribution for both Ni sites are almost the same and hence one can conclude that  the Ni ions have an average valence of +2.33 (\textit{d$^{7.67}$}), which leads to a maximum Ni moment of 2.33\,$\mu_B$. But in practice, oxides will not have a pure ionicity and often they have mixed bonding behavior i.e. one can expect noticeable covalent hybridization between the transition metal and surrounding atoms. As a result, the  the spin moment reduces owing to the fact that some of the electrons participate in bonding rather than in magnetism. So, transition$-$metal sites mostly have non-integer value for the spin moments and the surrounding atoms have small induced moments. Hence Ni1 closer to La is having a magnetic moment of 1.53\,$\mu_B$ and that closer to Sr has 1.58\,$\mu_B$ for the 75\% doped system. Similarly, Ni2 closer to La possesses a magnetic moment of 1.53\,$\mu_B$ and those closer to Sr has moment of 1.68\,$\mu_B$. This indicates that all the Ni atoms are in the HS state. Since La${_{4-x}}$Sr${_x}$Ni${_3}$O${_8}$ in practice will have oxygen deficiency at higher Sr substitution, one may expect a lower magnetic moments than the calculated values depending on the oxygen stoichiometry. Our calculated magnetic moments are valid only for the system without any oxygen vacancy.

\section{Summary}

We have made detailed investigation of the effect of hole-doping on electronic structure and magnetic properties of trilayer La${_4}$Ni${_3}$O${_8}$ using supercell approach. We have inferred the following:

\begin{enumerate}
\item Our calculations show that the \textit{C}$-$AFM state is the ground state with the Ni ions in the high spin state for La${_4}$Ni${_3}$O${_8}$, in agreement with experimental observations.
\item We have identified strong covalent interactions between Ni and O within the NiO${_2}$ square plane and ionic bonding between La and these building units in La${_4}$Ni${_3}$O${_8}$.
\item According to our results, La${_4}$Ni${_3}$O${_8}$ is a \textit{C}$-$AFM insulator with strong interlayer coupling. This insulating state  is caused by a strong Coulomb correlation effect and interlayer exchange coupling. Such strong interlayer coupling results from the high-spin occupation of Ni ions where the $d_{z^2}$ orbital gets occupied.
\item Our DOS calculation shows that in La${_4}$Ni${_3}$O${_8}$ both the Ni ions are in an average valence state.
\item Our GGA+\textit{U} calculations show intermediate band gap states originating from $d_{z^2}$ electrons from both the Ni ions which may be induced by the correlation effect.
\item The hole-doping brings insulator$-$to$-$metal transition without changing the \textit{C}$-$AFM ordering.
\item No signature of charge ordering is seen after the hole-doping which indicates that the average valency is sustained though the oxidation state for both the Ni atoms increases with the dopant concentration.
\item The magnetic moment gets enhanced in all the Ni sites due to the increase in the oxidation state of the Ni ions for the hole-doping cases.
\end{enumerate}

\section{Acknowledgments}
The authors are grateful to the Research Council of Norway for providing computing time at the Norwegian supercomputer facilities. This research was supported by the Indo-Norwegian Cooperative Program (INCP) via Grant No. F. No. 58-12/2014(IC) and Research Council of Norway, Grant No. 221905 (FRIPRO). One of the authors L.P. gratefully acknowledges the Central University of Tamil Nadu for the financial support.

\bibliography{LNO}

\end{document}